\documentclass[epj]{svjour}
\usepackage{graphics}
\usepackage{amsmath}
\usepackage{braket}
\usepackage{color}
\usepackage{xcolor}
\definecolor{dunkelgr}{cmyk}{1 0 1 0}
\usepackage{float}
\usepackage{bbm}
\usepackage{graphicx}
\usepackage{ulem}
\usepackage{hyperref}
\begin{document}
\title{Comparison of the Iterated Equation of Motion Approach and the 
Density Matrix Formalism for the Quantum Rabi Model}

\author{Mona Kalthoff\inst{1}, Frederik Keim\inst{1}, Holger Krull\inst{1} \and 
G\"otz S. Uhrig\inst{1}
}                     
%
%
\institute{\inst{1}Lehrstuhl f\"ur Theoretische Physik I, TU Dortmund, 
Otto-Hahn Stra\ss{}e 4, 44221 Dortmund, Germany}
\date{Received: 25 January 2017 / Received in final form 28 March 2017\\
The final publication is available at Springer via 
\href{http://rdcu.be/s2cT}{http://rdcu.be/s2cT}
}
%
\titlerunning{Comparison of the IEoM Approach and the DMF
}
\abstract{
The density matrix formalism and the equation of motion approach are two semi-analytical 
methods that can be used to compute the non-equilibrium dynamics of correlated systems. 
While for a bilinear Hamiltonian both formalisms yield the exact result, for any 
non-bilinear Hamiltonian a truncation is necessary. Due to the fact that the commonly 
used truncation schemes differ for these two methods, the accuracy of the obtained results 
depends significantly on the chosen approach. In this paper, both formalisms are applied to
the quantum Rabi model. This allows us to compare the approximate results and the exact
dynamics of the system and enables us to discuss the accuracy of the approximations as well as
the advantages and the disadvantages of both methods. It is shown to which extent the results fulfill physical requirements for the observables and which properties of the methods lead to 
unphysical results.  
} 
\maketitle
\section{Introduction}
\label{intro}

In about the last 15 years, refined experimental techniques based on ultracold 
gases in optical lattices, created by intense laser fields 
\cite{grein02b,kinos06,lewen07,bloch08,essli10},
have allowed for studies of closed quantum systems far away from equilibrium. 
Two facts are important: First, the systems must be well isolated in order not to
exchange energy with the environment quickly. In this way, long observation times are
possible. Second, the systems must be externally controllable
in time. This is achieved by manipulating the laser fields and other external electromagnetic fields. In this way, an externally controlled $H(t)$ can be tailored to the needs of
experiments which are not possible in solid state systems.
 One efficient way to push the system far out of equilibrium is to
start from an initial quantum state which is not an eigenstate of the Hamiltonian
$H(t\ge 0)$ which is constant for positive times. For instance, one may suddenly
change parameters which is called a quench.

Typically, the quenched systems are in highly excited states with respect
to the quenched Hamiltonian. Thus their dynamics is governed by processes 
on all energy scales including high energies. Properties 
may occur which are totally different from the equilibrium ones.
This makes the field of non-equilibrium physics 
particularly fascinating, both, from the
experimental and from the theoretical point of view.
While the earlier experiments dealt with bosonic system 
\cite{grein02b,kinos06,lewen07,bloch08,trotz12,ronzh13},
in the last years more and more investigations of fermionic systems
are performed \cite{stroh07,schne08,stroh10,essli10,schne12,perto14,cocch16}
or mixtures of both \cite{will15}. Recently, the spin degree of freedom
and its correlations are also addressed  
\cite{lubas11,kosch13,kraus14,eblin14,brown15,bohne16}.

The necessity to include all energy scales makes theoretical calculations, 
numerical or analytical ones, challenging \cite{eiser15}. The set of tools which can
be used is limited. So far, the majority of theoretical investigations were
focused on one-dimensional (1D) systems, on infinite dimensional
systems, and on small finite systems because for these cases powerful tools
are available. For 1D systems, the tool box is best: quantum field theoretical
descriptions provide analytical approaches
\cite{cazal06,uhrig09c,fiore10,sabio10,cazal11,schur12,rentr12}. 
The best understood models remain those which correspond to non-interacting 
fermionic or bosonic systems \cite{barth08,iucci10,calab12a,calab12b} 
or models which are effectively close to non-interacting ones \cite{kenne13}.
Time dependent density matrix renormalization group
is a powerful numerical tool which enables to study non-equilibrium phenomena
in 1D systems
\cite{daley04,white04a,schol05,schol11,manma07,karra12b,vidma13,zalet15,trotz12,perto14}.
The other dimensionality allowing for well-controlled studies 
is infinite dimensions where dynamical mean-field theory becomes exact
\cite{freer06,eckst09,schmi13,aoki14} 
and Gutzwiller approaches are  well justified \cite{schir10,schir11}.

Exact diagonalization is completely flexible concerning 
dimensionality, but it is restricted to small systems 
\cite{rigol07,rigol08,torre13}.
The intricate choice of basis states allows to reach even larger system size for specific issues \cite{bonca99,bonca07,mierz11a,mierz11b,bonca12}.
Recently, the technique of exact diagonalization has been boosted by using
it in a cluster approach for 1D systems \cite{rigol14}. 
While the results do not suffer from any finite size effects their validity is  limited 
by the maximum extension of the clusters which can be evaluated. 

A powerful macroscopic approach is to use the quantum Boltzmann equation
to describe the temporal evolution of the density. This works very well for
a variety of problems \cite{rapp10,schne12,eblin14}. Generically, the required
scattering matrix elements are taken from leading order perturbation theory.
Thus, complementary microscopic approaches are still desirable to verify the known
results and to extend them to large interactions.

So far, the question to which extent conserved quantities strongly restrict or even
prevent relaxation was in the center of interest
\cite{rigol07,fiore10,kolla11,polko11,calab12a,calab12b}.
Thus, integrable systems and systems close to integrability were studied, which drew the interest to 1D systems and to zero dimensional ones 
\cite{farib13a,farib13b,uhrig14a}.

Microscopic studies of two-dimensional (2D) models out of equilibrium are still very rare;
quantum Monte Carlo studies are an option \cite{goth12}. Other 2D studies address the 
influence of a strong electric field on the dynamics of carriers in a Mott insulator 
\cite{mierz11a,mierz11b,bonca12}. The life time of double occupancies in 2D models
 has been investigated experimentally and perturbatively \cite{stroh10}
and by exact diagonalization \cite{lenar13}.
Furthermore, the efficient representation of states as projected
entangled pairs is one of the promising numerical approaches to date
\cite{lubas11,schol11,orus14}. Still, the exponential growth of entanglement 
entropy restricts the application to short times in any dimension
\cite{gober05,eiser15} and poses a  constraint in particular in
two dimensions which is not easy to overcome. Of course, three dimensional systems
are even more difficult to describe reliably.

In solid state physics, pump-probe experiments provide a wealth of information
on the solid systems away from equilibrium 
\cite{rossi02,matsu12,matsu13b,aoki14,matsu16,ramea16}. 
Currently, particular interest is devoted to stirring ordered phases such as
charge density waves \cite{perfe06,schmi08e} or superconducting phases 
\cite{matsu12,matsu13b,matsu16,akbar13,krull14,kempe15,krull16}.

The above brief review, which cannot be exhaustive, illustrates impressively that
the field of non-equilibrium physics in general is currently an extremely active field of research. Thus, the development of theoretical approaches and their assessment
is a timely task. To this end, we study two approaches based on the Heisenberg equations
of motion and compare them in the present article. The two approaches are the density-matrix
formalism (DMF) and the iterated equations of motion (iEoM), see below.
They will be applied to the quantum Rabi model (QRM) which is overseeable enough to 
understand the origin of the observed behavior. In addition, its great advantage
is that an exact solution exists which we can use to gauge the approximate approaches.
Thus, the QRM provides an ideal testbed. Our study is intended to render the
application of either of the approximate approaches to more complex extended models
more efficiently.

The article is set up as follows. In Sect.\ \ref{sec:rabi}, we briefly introduce the model
which serves as the testbed. In Sect.\ \ref{sec:methods}, the two approximate approaches
are introduced and their applications to the model are explained. The results are presented
in Sect.\ \ref{sec:results}. Finally, our findings are summarized in the Conclusions
\ref{sec:conclusio} where we provide an outlook as well.

\section{The quantum Rabi model}
\label{sec:rabi}

The quantum Rabi model (QRM), which was introduced in 1936 by Rabi \cite{rabi36}, describes the interaction between a single bosonic mode and a two-level system. It is one of the simplest strongly coupled quantum systems. The Hamiltonian reads 
\begin{equation}
\label{eqn:HamiltonianRabi}
H_R=\underbrace{\frac{\omega_0}{2}\sigma_z}_{H_\sigma} +
\underbrace{\omega b^\dagger b}_{H_B}
+ \underbrace{g\sigma_x \left(b^\dagger +b\right)}_{H_I} ,
\end{equation}
where $b^\dagger$ and $b$ are the bosonic creation and annihilation operators, $\omega$ 
 the boson frequency and $\sigma_x$ and $\sigma_z$ are Pauli matrices describing the two-level
 system. The two-level system represented by $H_\sigma$ is characterized by
 the energy difference $\omega_0$. It is coupled to the bosonic environment $H_B$. The 
coupling between system and environment is described by the interaction Hamiltonian $H_I$
with coupling parameter $g$. 

Every state of the Rabi model can be expanded in product states of a bosonic state and 
a spin state. In this basis, both $H_\sigma$ and $H_B$ can be chosen diagonal. 
However, there is no common eigenbasis for $\sigma_x$ and $\sigma_z$. 
Therefore, there is no common eigenbasis of all parts of the
Hamiltonian. Only the total energy is preserved. 
We choose $\omega$ as our energy unit, i.e., all energies are given in units of $\omega$. 
Also, $\hbar$ is set to unity for convenience. 

One key advantage of this model is that an exact solution is available. On the basis of discrete symmetries, Braak obtained an exact set of eigenstates and eigenvalues in 2011 
\cite{braak11}. On this basis the numerically exact time evolution of the observables 
$\braket{\sigma_x}$ and $\braket{\sigma_z}$ \cite{wolf12} was computed.
Another advantage of the QRM is, that the expectation values of the spin operators as well as of the number operator have to comply with certain constraints. Expectation values such as 
$\braket{b^\dagger b}$ can be rewritten according to
\begin{equation}
\label{eqn:nichtnegativ}
\bra{\Psi(t)}b^\dagger b\ket{\Psi(t)}=\left|b\ket{\Psi(t)}\right|^2 .
\end{equation} 
Therefore the expectation value of the number operator remains non-negative at all times.

Every two-level system can be represented by a spin $S=\frac{1}{2}$. A maximum polarization 
in one direction corresponds to an expectation value 1 of the corresponding Pauli matrix.
The expectation values of the Pauli matrices are bounded between -1 and 1. 
The number operator, the spin operators and the energy are self-adjoint operators, so their expectation values are real.

\section{Methods}
\label{sec:methods}

Here we briefly present the two general approaches used and compared in this
article. 

\subsection{The density matrix formalism}
\label{sec:DMF}

The density matrix formalism (DMF) derives the equations of motion of expectation values by 
using the Heisenberg equation of motion. If the considered operators are not explicitly time 
dependent, but the time dependence is only induced by the Hamiltonian, the Heisenberg equation 
of motion reads
\begin{equation}
\label{eqn:heisenbergschebwgl}
\frac{\mathrm{d}}{\mathrm{d}t}A=-i\left[A,H\right] .
\end{equation}    
In the Heisenberg picture, the states $\ket{\Psi}$ and the corresponding density matrices 
$\rho_\Psi$ are time independent. Therefore the temporal evolution of an expectation value is 
given by 
\begin{subequations}\label{eqn:dichtegl}
\begin{eqnarray}
\frac{\mathrm{d}}{\mathrm{d}t} \braket{A}_\Psi &=& 
\frac{\mathrm{d}}{\mathrm{d}t}\mathrm{Tr}\left(\rho_\Psi A\right)
= \mathrm{Tr}\left(\rho_\Psi \frac{\mathrm{d}}{\mathrm{d}t}A\right)
\\ 
&=& -i\mathrm{Tr}\left(\rho_\Psi\left[A,H\right]\right)=
-i\braket{\left[A,H\right]}_\Psi .
\end{eqnarray}
\end{subequations}
Depending on the Hamiltonian, calculating the commutator of the operator $A$ and the Hamiltonian may (and generically will) result in the appearance of additional 
expectation values whose temporal evolutions have to be calculated again by applying equation 
\eqref{eqn:dichtegl}. 
In case of a bilinear fermionic or bosonic Hamiltonian, this procedure results in a closed set
 of differential equations that can be computed exactly. But in the case of interacting
Hamiltonians, for instance the QRM, the application of \eqref{eqn:dichtegl} 
leads to an infinite hierarchy of differential equations which does not close.
Hence, these equations cannot be integrated straightforwardly.

A truncation is necessary in order to obtain a closed system so that the temporal 
evolution of expectation values can be computed. 
One way to truncate is to neglect the interaction between operators if the corresponding  cumulant exceeds a certain order. The order of the cumulant is given by the number of
operators appearing in it \cite{kubo62,fulde88}.
The fundamental idea of the DMF is that the contribution of the cumulant including a larger number of operators, i.e., being of higher order, is smaller.

Cumulants occur upon factorizing the expectation value of an operator product \cite{kubo62}. 
They  are calculated according to  
\begin{equation}
\label{eqn:cumulant}
\braket{A_1^\alpha ... A_n^\beta}^c=
\left.\frac{\partial^\alpha}{\partial \lambda_1^\alpha}...\frac{\partial^\beta}{\partial \lambda_n^\beta}
\ln\left\langle
\prod_{i=1}^n e^{\lambda_i A_i}\right\rangle
\right|_{\lambda_1= ...=\lambda_n=0}
\end{equation} 
and comprise the dynamics of the interaction between the operators in the initial product. 
For instance, the cumulants for expectation values consisting of one, two and three operators 
are given by 
\begin{subequations}
\begin{eqnarray}\label{eqn:Kumulanten}
\braket{A}^c&=&\braket{A}
\label{eqn:Kumulanten1}
\\
\braket{AB}^c&=&\braket{AB}-\braket{A}\braket{B}
\label{eqn:Kumulanten2}
\\\nonumber
\braket{ABC}^c&=&\braket{ABC}+2\braket{A}\braket{B}\braket{C}
\label{eqn:Kumulanten3}\\
&&-\braket{A}\braket{BC}- 
\braket{B}\braket{AC} -
\braket{C}\braket{AB} .
\end{eqnarray}
\end{subequations}
The number $n$ in \eqref{eqn:cumulant} denotes the order of the cumulant, i.e.,
 the number of operators occuring in it. 

To construct the truncation by cumulants, the system of differential equations for the expectation values has to be converted into a system of differential equations 
for the corresponding cumulants. For instance, the temporal evolution of 
$\braket{\sigma_z b}$ can be expressed in cumulants according to
\begin{subequations}
\begin{eqnarray}
\nonumber\frac{\mathrm{d}}{\mathrm{d}t} \braket{\sigma_z b} &=&
-i\omega \braket{\sigma_z b}
+g\braket{\sigma_y}\\
&&+2g\braket{\sigma_y b b}
+2g\braket{\sigma_y b^\dagger b}\\\nonumber
&=&-i\omega \braket{\sigma_z b}^c -i\omega \braket{\sigma_z} \braket{b}
+g\braket{\sigma_y}\\\nonumber
&&+2g\left(\braket{\sigma_y b}^c\braket{b}
+\braket{\sigma_y b}^c\braket{b} + \braket{b b}^c\braket{\sigma_y}
\right)\\\nonumber
&&+2g\left( \braket{\sigma_y b}^c\braket{b^\dagger}
+\braket{\sigma_y b^\dagger}^c\braket{b} + \braket{b^\dagger b}^c\braket{\sigma_y}
\right)\\\nonumber
&&+2g\left(\braket{\sigma_y} \braket{b}\braket{b}+\braket{\sigma_y} 
\braket{b^\dagger}\braket{b}\right)\\
&&+2g\left({\braket{\sigma_y b^\dagger b}^c}+\braket{\sigma_y b b}^c\right) .
\label{eqn:kumulantenbeispiela}
\end{eqnarray} 
\end{subequations}

Furthermore, using the product rule for calculating the derivative with respect to the 
time leads to
\begin{subequations}
\begin{eqnarray}
\nonumber\frac{\mathrm{d}}{\mathrm{d}t}\braket{\sigma_z b} &=& 
\left(\frac{\mathrm{d}}{\mathrm{d}t}\braket{\sigma_z b}^c\right)\\
&& + \braket{\sigma_z}^c\left(\frac{\mathrm{d}}{\mathrm{d}t} \braket{b}^c\right)
 + \braket{b}^c\left(\frac{\mathrm{d}}{\mathrm{d}t} \braket{\sigma_z}^c\right)
\\
&=&\nonumber\left(\frac{\mathrm{d}}{\mathrm{d}t}\braket{\sigma_z b}^c\right)- 
i\omega \braket{b}\braket{\sigma_z} -ig\braket{\sigma_x}\braket{\sigma_z}
\\
\nonumber &&+ 2g\left(\braket{\sigma_y b}^c\braket{b}+\braket{\sigma_y b^\dagger}^c
\braket{b} \right)
\\
&&+2g\left(\braket{\sigma_y}\braket{b}\braket{b}+
\braket{\sigma_y}\braket{b^\dagger}\braket{b}\right) .
\label{eqn:kumulantenbeispielb}
\end{eqnarray}
\end{subequations}
Equating \eqref{eqn:kumulantenbeispiela} and \eqref{eqn:kumulantenbeispielb} 
yields the differential equation for the cumulant
\begin{eqnarray}
\frac{\mathrm{d}}{\mathrm{d}t}\braket{\sigma_z b}^c=
 &-& i\omega \braket{\sigma_z b}^c -ig\braket{\sigma_x}\braket{\sigma_z}
\\\nonumber
 &+&g\braket{\sigma_y}\\\nonumber
 &+& 2g\left({\color{red}\braket{\sigma_y bb}^c} + \braket{\sigma_y b}^c\braket{b} + 
\braket{ bb}^c\braket{\sigma_y} \right)
\\\nonumber
  &+& 2g\left({\color{red}\braket{\sigma_y b^\dagger b}^c} + 
	\braket{\sigma_y b}^c\braket{b^\dagger} + 
	\braket{ b^\dagger b}^c\braket{\sigma_y} \right) .
\end{eqnarray} 
The cumulants marked in red ({\color{red}{dark grey}}) are of third order because the number of operators involved 
in the cumulant is three. Thus, these cumulants are truncated in an approximation of 
second order. We emphasize that this is not equivalent to a perturbative truncation in 
the order of the coupling parameter $g$. 
The truncation based on the order of cumulants only requires the assumption
 that the contribution of a cumulant decreases with the number of operators involved. 
This leads to a closed system of differential equations amenable to numerical solution.

\subsection{The iterated equation of motion approach}
\label{sec:ieom}

Similar to the DMF, the approach of  iterated equations of motion (iEoM) is based on the 
Heisenberg equation of motion. But instead of directly deriving the temporal evolution of
 certain expectation values, the temporal evolution of certain operators is computed 
\cite{uhrig09c,hamer13a,hamer13b,hamer14a,krull15a}. As in the DMF, the commutator in the Heisenberg equation of motion leads to an infinite hierarchy of differential equations 
for generic Hamiltonians.  Thus, in order to solve for the temporal evolution of the 
operators, a truncation becomes necessary. To this end, we start from an operator basis
for which we solve the iEoM. The basic idea is that the approximation becomes
more and more accurate upon extending this basis.

Such an operator basis can be obtained by systematically 
extending the ansatz for each initial operator
or by defining a common operator basis for all 
operators under study. The time dependences of the operators in the Heisenberg 
picture being elements of the chosen basis are put into prefactors $\gamma_{nl}(t)$ 
of the time independent operators $V_n^S$ in the Schr\"odinger picture. Hence, the ansatz 
for the dynamics of an operator $V_n^H$ in the Heisenberg picture is given by
\begin{equation}
\label{eqn:ansatz}
V_i^H(t)=\sum_l \gamma_{nl}(t) V_l^S .
\end{equation}     
The initial conditions of the prefactors are 
\begin{equation}
\gamma_{nl}(t=0)=
\begin{cases}
  1  & \text{for }l=n\\
  0 & \text{for }l\neq n
\end{cases} .
\end{equation} 

As mentioned above, the operator basis can be obtained by initially only considering the operator $V_n$ and extending the basis systematically by operators $V_\text{new}$ 
occurring in the commutator in Eq.~\eqref{eqn:heisenbergschebwgl}. 
The resulting differential equations can be truncated strictly in the order of the 
coupling parameter $g$ by including only operators of $\mathcal{O}(g^n)$ in the operator 
basis.

The alternative approach is to define a common operator 
basis for all operators. If such a common operator basis is defined beforehand
the truncation is not applied in a strict order of the coupling parameter $g$.
Contributions of  $\mathcal{O}(g^{n+m})$ are included if the corresponding operator 
is an element of the chosen basis. Thus, the results obtained by truncating using an 
operator basis will differ from the results obtained by truncating strictly in the order
of the coupling parameter. 

For instance, the initial ansatz for the annihilation operator is given by 
\begin{equation}
\label{eqn:ansatz0}
b^H(t)=\beta_0(t)b^S .
\end{equation}
Heisenberg's equation of motion implies
\begin{equation}
\frac{\mathrm{d}}{\mathrm{d}t} b(t)=\beta_0\left(-i\left[b,H_R\right]\right)=
\beta_0\left(-i\omega b-ig\sigma_x\right) ,
\end{equation}
so $\sigma_x$ is included in the extended ansatz which now reads 
\begin{equation}
b(t)=\beta_0(t)b+g\beta_1(t)\sigma_x .
\end{equation}
This can again be computed using the Heisenberg equation of motion. Extending the ansatz 
systematically and only considering contributions in $\mathcal{O}(g^1)$
 eventually leads to 
\begin{equation}
\label{eqn:ansatz2}
b(t)=\beta_0(t)b+g\beta_1(t)\sigma_x+g\beta_2(t)\sigma_y ,
\end{equation}
so that the temporal evolution of the annihilation operator is given both by
\begin{subequations}
\begin{eqnarray}
\frac{\mathrm{d}}{\mathrm{d}t} b(t)&=&
\beta_0\left(-i\omega b-ig\sigma_x\right)+
g\beta_1 \left(-\omega_0\sigma_y\right)\\\nonumber &&+
g\beta_2 \left(\omega_0\sigma_x\right)\\
&=&\label{eqn:weg1}b\left(-i\omega\beta_0\right)
+g\sigma_x\left(-i\beta_0+\omega_0\beta_2\right)\\\nonumber &&+
g\sigma_y\left(-\omega_0\beta_1\right)
\end{eqnarray}
\end{subequations}
and by
\begin{equation}
\label{eqn:weg2}
\frac{\mathrm{d}}{\mathrm{d}t} b(t)=
b\frac{\mathrm{d}}{\mathrm{d}t}\beta_0+
g\sigma_x\frac{\mathrm{d}}{\mathrm{d}t}\beta_1+
g\sigma_y\frac{\mathrm{d}}{\mathrm{d}t}\beta_2 .
\end{equation}

Comparing Eqs.\ \eqref{eqn:weg1} and \eqref{eqn:weg2} yields the closed system of 
differential equations 
\begin{subequations}
\label{eqn:dglb}
\begin{eqnarray}
\frac{\mathrm{d}}{\mathrm{d}t}\beta_0&=&-i\omega\beta_0\\
\frac{\mathrm{d}}{\mathrm{d}t}\beta_1&=&-i\beta_0+\omega_0\beta_2\\
\frac{\mathrm{d}}{\mathrm{d}t}\beta_2&=&
-\omega_0\beta_1 ,
\end{eqnarray}
\end{subequations}
which can be solved numerically. In this case, truncating strictly in the coupling parameter 
$g$ and truncating according to the operator basis yields the same result. 
But if contributions in $\mathcal{O}(g^2)$ are considered, 
computing Heisenberg's equation of motion for the ansatz 
\begin{eqnarray}
\label{eqn:ansatz2ord}
b(t)=\beta_0(t)b+g\beta_1(t)\sigma_x+g\beta_2(t){\color{dunkelgr}{\sigma_y}}
\\\nonumber
+g^2\beta_3(t)\sigma_z b
+g^2\beta_4(t)\sigma_z b^\dagger
\end{eqnarray}
leads to 
\begin{eqnarray}
\label{eqn:iEoMvernichter2ord}
\frac{\mathrm{d}}{\mathrm{d}t} b(t)&=&
\beta_0\left(-i\omega b-ig\sigma_x\right)
\\\nonumber
&&+ g\beta_1 \left(-\omega_0\sigma_y\right)+
g\beta_2 \left(\omega_0\sigma_x-2g\sigma_z b-2g\sigma_z b^\dagger\right)
\\\nonumber 
&&+g^2\beta_3 \left(
-i\omega\sigma_z b + {\color{dunkelgr}{g\sigma_y}}+
{\color{red}{2g\sigma_y bb +2g\sigma_y b^\dagger b }}\right)
\\\nonumber
&&+g^2\beta_4 \left(i\omega\sigma_z b^\dagger + {\color{dunkelgr}{g\sigma_y}}+
{\color{red}{2g\sigma_y b^\dagger b^\dagger +2g\sigma_y b^\dagger b }}\right) .
\end{eqnarray}

In a strict truncation in $\mathcal{O}(g^2)$, both the operators highlighted in
green ({\color{dunkelgr}{light grey}}) and in red ({\color{red}{dark grey}}) are neglected. This differs from the truncation in the more general 
approach based on a pre-defined operator basis. Using the pre-defined operator basis
 the green ({\color{dunkelgr}{light grey}}) contributions in $\mathcal{O}(g^3)$ are kept because $\sigma_y$,  which can be found in the ansatz \eqref{eqn:ansatz2ord}, is an element of the operator basis.

The definition of a common operator basis for all operators allows us to describe 
the dynamics of the operators in matrix notation. 
The temporal evolution of an operator that is not explicitly time dependent is given by 
\begin{equation}
\label{eqn:Matrix}
\frac{\mathrm{d}}{\mathrm{d}t}V_n= i\left[V_n,H\right] = i\sum_{j} M_{nj}V_{j}.
\end{equation}
The dynamics of the vector $\vec{V}$ containing all operators in the basis can therefore be described using the matrix $\uuline{M}$ according to  
\begin{equation}
\label{eqn:matrixdgl1}
\frac{\mathrm{d}}{\mathrm{d}t}\vec{V}^H = i\uuline{M}\vec{V}^H .
\end{equation}
Henceforth, we will call the transposed matrix $\uuline{M}^T$ the dynamic matrix belonging
to the chosen basis of operators.
The Hamiltonian $H_R$ is not time dependent and thus equal in both the Schr\"odinger 
and the Heisenberg picture. The Heisenberg equation of motion can be written as  
\begin{subequations}
\begin{eqnarray}
i\left[V_n^H,H\right]&=&-i\left[\sum_{j}\gamma_{nj}V_{j}^S,H\right]\\
&=&\sum_{j}\gamma_{nj}\left(-i\left[V_{j}^S,H\right]\right)\\
&=&i\sum_{j}\gamma_{nj}
\sum_{k} M_{jk}V_{k}^S\\
&=&i\sum_{k}V_{k}^S\sum_{j}
\gamma_{nj}M_{jk} .
\label{eqn:vdot1}
\end{eqnarray}
\end{subequations}

Using \eqref{eqn:ansatz}, the commutator yields
\begin{eqnarray}
\label{eqn:vdot2}
i\left[V^H_n,H\right]= \frac{\mathrm{d}}{\mathrm{d}t}V^H_n
=\sum_{k} V_{k}^S\frac{\mathrm{d}}{\mathrm{d}t}\gamma_{nk} .
\end{eqnarray} 
Comparing Eqs.\ \eqref{eqn:vdot1} and \eqref{eqn:vdot2} we obtain
\begin{subequations}
\begin{eqnarray}
\frac{\mathrm{d}}{\mathrm{d}t}\gamma_{nk}&=&i\sum_{j}\gamma_{nj}
M_{jk}\\
\frac{\mathrm{d}}{\mathrm{d}t}\uuline{\gamma}&=&i\uuline{\gamma}\uuline{M}
,
\label{eqn:Matrixgleichung}
\end{eqnarray}
\end{subequations}
where the latter notation underlines the matrix form of the equation. 
Both $\uuline{\gamma}$ and $\uuline{M}$ are quadratic matrices. 
Eq.\ \eqref{eqn:Matrixgleichung} is equivalent to the vector equation 
\begin{equation}
\label{eqn:trans}
\frac{\mathrm{d}}{\mathrm{d}t}\left(\vec{\gamma}_n\right)^T
=i\left(\vec{\gamma}_n\right)^T \uuline{M} .
\end{equation}
Here, $\vec{\gamma}_n$ is a column vector made from the elements of the $n$-th 
\textit{row} of the matrix $\uuline{\gamma}$. Transposing \eqref{eqn:trans} leads to the vectorial differential equation 
\begin{equation}
\label{eqn:matrixdgl2}
\frac{\mathrm{d}}{\mathrm{d}t}\vec{\gamma}_n=i\uuline{M}^T \vec{\gamma}_n ,
\end{equation}
that is solved by 
\begin{equation}
\label{eqn:loesungkoeff}
\vec{\gamma}_n=\sum_{m} C_{nm} e^{i\lambda_{m} t} \vec{u}_{m} .
\end{equation}
Here, $\lambda_{m}$ denotes the eigenvalues and $\vec{u}_{m}$ the eigenvectors of the 
dynamic matrix $\uuline{M}^T$. Therefore, choosing an operator basis that corresponds to a 
\textit{hermitian} matrix $\uuline{M}$ guarantees that the temporal evolution
consists of oscillatory contributions only as is physically reasonable in
quantum mechanics. No exponentially diverging time dependence
can occur. Conversely, if the dynamic matrix $\uuline{M}^T$ is not hermitian complex eigenvalues
may appear which imply exponentially diverging contributions which are unphysical.

For the QRM no operator basis can be found that implies a hermitian dynamic matrix 
$\uuline{M}^T$ and comprises all operators in $\mathcal{O}(g^2)$. 
But there is a basis that leads to a 
matrix with only real eigenvalues satisfying exactness in $\mathcal{O}(g^3)$. 
This basis will be discussed more specifically in the following section.

We stress an inherent advantage of the iEoM approach, namely that the
 non-negativity of the number operator is guaranteed. Calculating the temporal evolution of 
$b(t)$ the number operator can be expressed by $b^\dagger(t)b(t)$ which is 
a non-negative operator by construction. Thus, the resulting expectation values will always
 be non-negative as it has to be on physical grounds.

\section{Results}
\label{sec:results}

The systems of differential equations for both the DMF and the iEoM approach are solved numerically using the Livermore Solver for Ordinary Differential Equations \cite{radha93} or, alternatively, a Runge Kutta algorithm of fourth 
order~\cite{dorma86}.  The system of differential equations for the DMF cannot be 
integrated for long time intervals with either of these algorithms because the numerically iterated matrix of coefficients becomes singular within machine precision.

The energy splitting of the spin $\omega_0$ used below in the computations of the 
temporal evolutions of the expectation values is 
$\omega_0=0.5\omega$. This value is chosen because it was also used
in the exact data provided in Ref.~\cite{wolf12}. For the iEoM, two different operator bases are considered. Both comprise the elementary set of operators
\begin{eqnarray}
\label{eqn:Operatorbasis_vorfaktoren}
V &:=&\left\lbrace \sigma_x,\,\sigma_y,\,\sigma_z\right\rbrace
\\ \nonumber
&&\cup
\left\lbrace\sqrt{2}\sigma_x b,\,
\sqrt{2}\sigma_y b,\,
\sqrt{2}\sigma_z b\,\right\rbrace 
\\\nonumber
&&\cup
\left\lbrace\sqrt{2}\sigma_x b^\dagger,\,
\sqrt{2}\sigma_y b^\dagger,\,
\sqrt{2}\sigma_z b^\dagger\right\rbrace 
\\\nonumber
&&\cup \left\lbrace \sqrt{2}\sigma_x bb,\,
\sqrt{2}\sigma_y bb,\,\sqrt{2}\sigma_z bb,\,
 \right\rbrace\\\nonumber
&&\cup \left\lbrace \sqrt{2} \sigma_x b^\dagger b^\dagger ,\, \sqrt{2}\sigma_y 
b^\dagger b^\dagger,\,
\sqrt{2}\sigma_z b^\dagger b^\dagger \right\rbrace\\\nonumber
&&\cup
\left\lbrace 
 2\sigma_x b^\dagger b,\, 2\sigma_y b^\dagger b,\, 2\sigma_z b^\dagger b \right\rbrace\,,
\end{eqnarray}
where the pre-factors are chosen such that the ensuing dynamic matrix $\uuline{M}^T$ 
is hermitian.
However, the basis $V$ is not sufficient to describe the QRM because  $H_B$ in 
\eqref{eqn:HamiltonianRabi} contains the bosonic operators without spin operators attached
but neither $b^\dagger b$, nor $b$ or $b^\dagger$ are elements of $V$.
Thus, we consider two extensions
\begin{subequations}
\begin{eqnarray}
V_{\mathrm{sep}} &:=& V
\cup \left\lbrace \mathbbm{1},\,b,\,b^\dagger \right\rbrace\\
V_{\mathrm{prod}} &:=& V
\cup \left\lbrace \mathbbm{1},\, \sqrt{2} b^\dagger b \right\rbrace.
\end{eqnarray}
\end{subequations}
The difference between both bases is that one ($V_{\mathrm{prod}}$) contains $b^\dagger b$ in
 a product while the other ($V_{\mathrm{sep}}$) 
contains the bosonic creation and annihilation operator separately. 

While none of the two operator bases leads to a hermitian dynamic matrix $\uuline{M}^T$, 
we find that  the matrix resulting from $V_{\mathrm{prod}}$ only has real eigenvalues. 
Therefore,  no exponential divergences occur if this basis is used to calculate the 
expectation values.
In contrast, the matrix resulting from $V_{\mathrm{sep}}$ has complex eigenvalues occurring 
in conjugate pairs with finite imaginary parts. This implies that exponentially diverging solutions for the expectation values occur. 
Independent from the divergence behavior the expectation value 
$\langle b^\dagger b\rangle(t)$ remains non-negative
at all times.

\subsection{Temporal evolution of the spin operators}
\label{sec:so}

\begin{figure}
 \centering
   \includegraphics[width=\columnwidth]{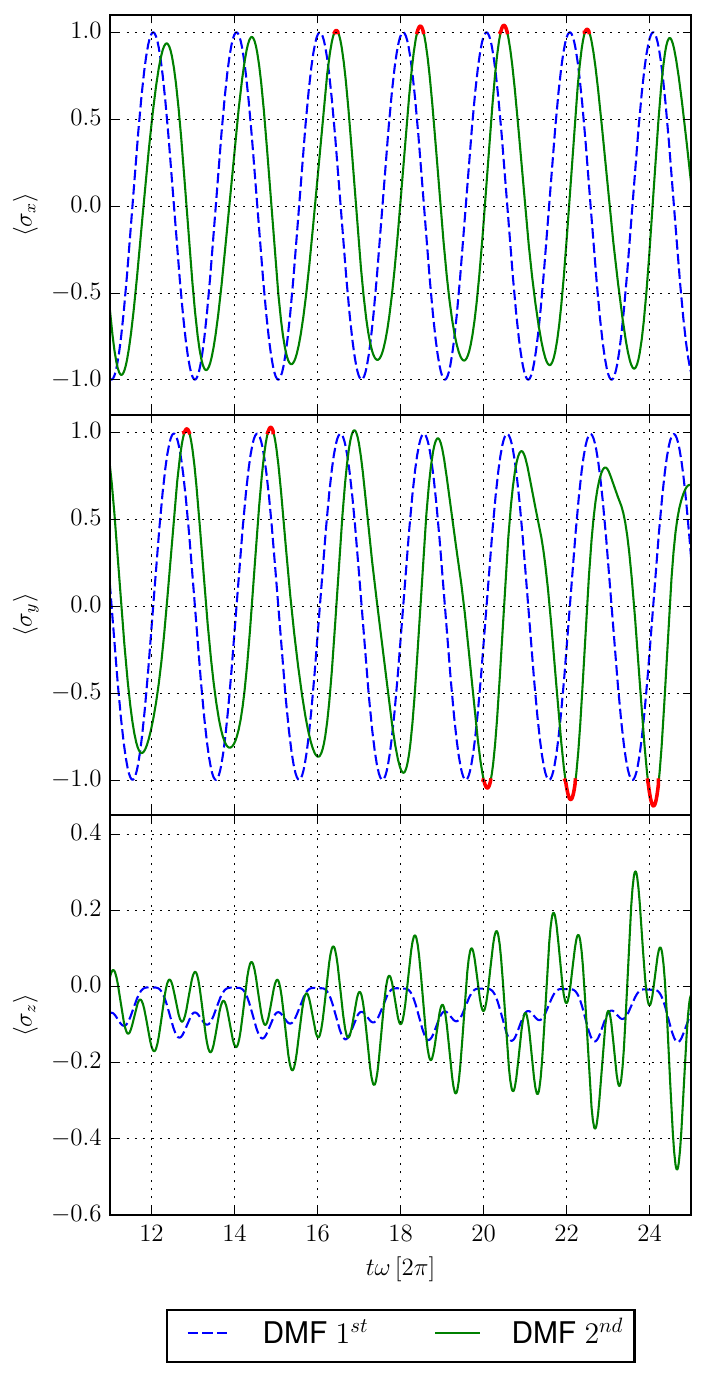} 
   \caption{DMF results for the temporal evolution of the expectation values of the
	spin operators starting from
	$\braket{\sigma_x}(0)=1$, $\omega_0=0.5\omega$, and $g=0.1\omega$ in first and second order. Expectation values with $|\braket{\sigma}|>1$ in second order are marked in red.}
  \label{f:spinoperatoren_DMF}
\end{figure} 

\begin{figure}
 \centering
   \includegraphics[width=\columnwidth]{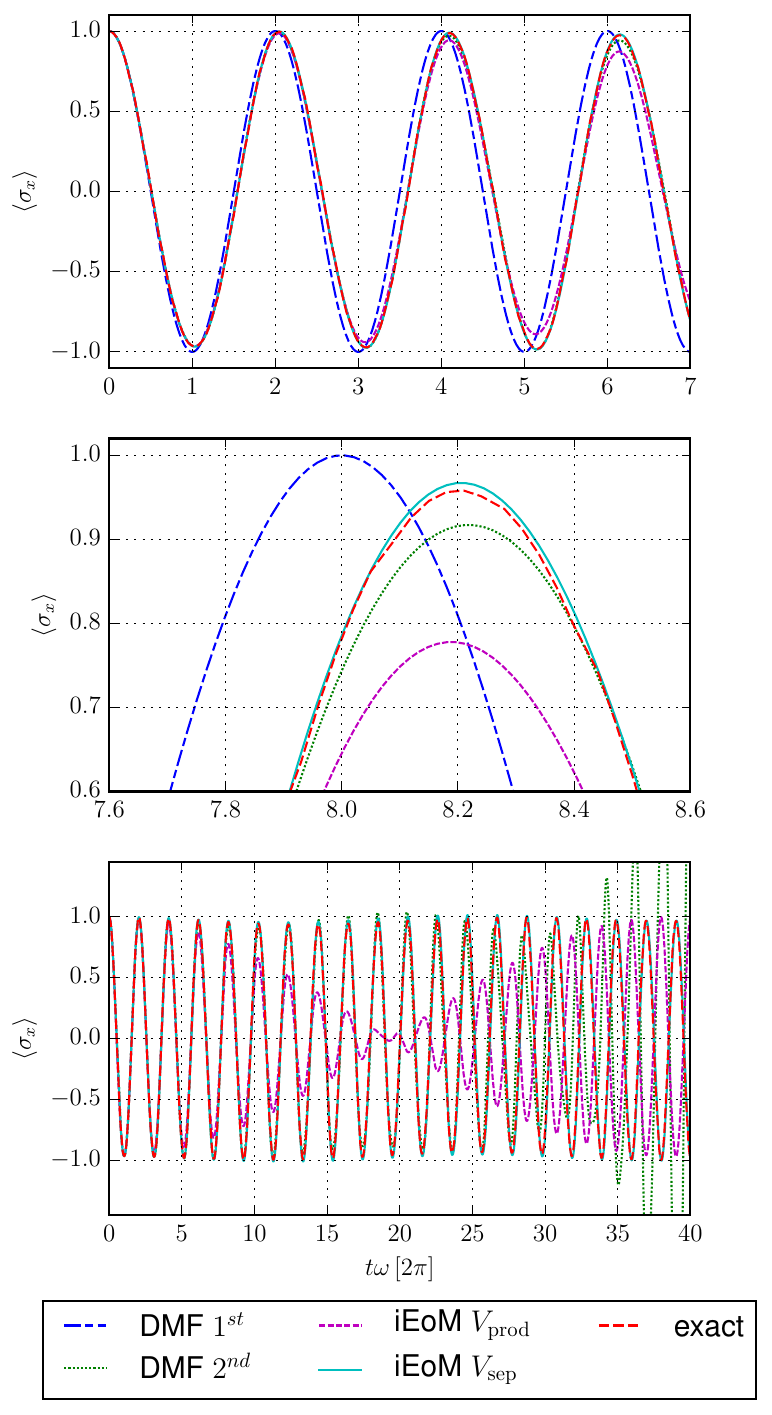}   
   \caption{Results for the temporal evolution of $\braket{\sigma_x}$  starting from
	$\braket{\sigma_x}(0)=1$, $\omega_0=0.5\omega$, and $g=0.1\omega$, computed using the DMF 
	in first and second order as well as the iEoM for $V_{\mathrm{prod}}$ and 
	$V_{\mathrm{sep}}$. To show the agreement with the exact solution, three 
	different time ranges are displayed. Because the first order results in the DFM differ 
	significantly from the exact solution even on short time scales, 
	they are not displayed in the lowest panel.}
  \label{f:sx_vergleich}
\end{figure} 

Due to the truncation, the DMF approach may lead to unphysical behaviour, for instance the 
expectation values for the spin operators can have absolute values beyond unity or the
expectation value for the particle number turns negative.
The QRM is an excellent testbed to study and to analyze such behavior. 
Fig.~\ref{f:spinoperatoren_DMF} shows the DMF results for the temporal evolution of the spin operators in first and second order for $g=0.1\omega$, where the initial state is set to
\begin{equation}
\ket{\Psi_0}=\ket{\chi}\otimes\ket{0} ,
\end{equation}
that is to say that the spin is polarized in $x$-direction and the bosonic mode is 
initially unoccupied. In first order, where the iEoM and the DMF approach produce the same 
results, $\braket{\sigma_x}$ and $\braket{\sigma_y}$ oscillate with a constant phase shift 
of $\frac{\pi}{2}$. This corresponds to a rotation in the $xy$-plane as expected for a 
precessing spin (Larmor precession). 
Inspecting the second order results of the DMF, it becomes evident that 
for $t\omega[2\pi] > 12.5$ the absolute expectation values of the Pauli matrices
exceed unity which clearly is unphysical.

Fig.~\ref{f:sx_vergleich} displays the results of the DMF in comparison to the exact solution
 from Ref.~\cite{wolf12} and to the results of the iEoM using the operator bases 
$V_{\mathrm{prod}}$ and $V_{\mathrm{sep}}$. It should be noted that calculating 
the temporal evolution of the spin operators using the iEoM, a strict truncation in 
second order of the coupling parameter equals a truncation using the operator basis 
$V_{\mathrm{prod}}$. 

Looking at short times, it is obvious that the DMF results in first order differ significantly
 from the exact solution in both amplitude and phase.
But the second order DMF results  agree better with the exact solution than the results of the
 iEoM using $V_{\mathrm{prod}}$. At the same time, the temporal evolution computed using the iEoM based on $V_{\mathrm{sep}}$ is in remarkable agreement with the exact evolution. To 
illustrate this good agreement, a zoom of the peak of the temporal evolution around 
$t\omega=2\pi\cdot 8.2$ is displayed in the middle panel in Fig.~\ref{f:sx_vergleich}. 
This excellent agreement persists for longer times as displayed in the lower panel. 
In addition, the lower panel shows that the expectation value computed using second order
 DMF diverges exponentially which is a serious caveat physically.
In contrast, the results of the iEoM using $V_{\mathrm{prod}}$ show no unphysical behavior 
and no exponential divergences. This remains true for larger coupling parameters and was to be expected because the dynamic matrix $\uuline{M}^T$ resulting from $V_{\mathrm{prod}}$ only 
has real  eigenvalues.

But the iEoM approach based on $V_{\mathrm{prod}}$ leads to large discrepancies to the exact solution. It is remarkable that the 
analytically better justified ansatz leads to qualitatively worse results. 
While the computation using $V_{\mathrm{sep}}$ implying complex eigenvalues agrees 
significantly better, the expectation values of the spin operators show unphysical results 
exceeding unity, even though no divergences occur on the considered time scales.
Because the deviations of the results computed using the DMF in $\text{1}^\text{st}$ order become evident on short time scales, these results are not displayed for larger time scales in the lowest panel.

We depict only the temporal evolution of $\braket{\sigma_x}$, but the computation of 
$\braket{\sigma_z}$ yields similar results when using $V_{\mathrm{prod}}$ and 
$V_{\mathrm{sep}}$. 
This is due to the block structure of the dynamic matrix $\uuline{M}^T$ 
in \eqref{eqn:matrixdgl2}.
There are submatrices which are not linked to the other blocks so that the
dynamics of certain operators is not coupled to the other operators in the operator basis.
This means that for certain initial conditions reduced bases $V_\mathrm{red}$ 
can be identified yielding the same temporal evolution as the compuation in the full basis. 

In concrete terms, neither the coefficients corresponding to $b$ and $b^\dagger$ nor the coefficient belonging to the number operator appear in the evolution of $\sigma_z$. 
Hence, the reduced basis
\begin{eqnarray}
\label{eqn:Operatorbasis_vorfaktoren_red}
V_\mathrm{red} &:=&\left\lbrace 
\sigma_z,\,\mathbbm{1}
\right\rbrace\\&&\nonumber
\cup
\left\lbrace
\sqrt{2}\sigma_x b,\,
\sqrt{2}\sigma_x b^\dagger
\,\right\rbrace 
\\\nonumber
&&\cup
\left\lbrace
\sqrt{2}\sigma_y b,\,
\sqrt{2}\sigma_y b^\dagger
\right\rbrace 
\\\nonumber
&&\cup \left\lbrace\sqrt{2}\sigma_z bb,\,
\sqrt{2}\sigma_z b^\dagger b^\dagger ,\,
 2\sigma_z b^\dagger b
 \right\rbrace\\\nonumber
\end{eqnarray}
is sufficient and the corresponding dynamic matrix is hermitian.
Therefore, the results for the temporal evolution of $\braket{\sigma_z}$ agree 
almost perfectly with the exact temporal evolution for small coupling parameters
and even for larger coupling parameters no unphysical behavior is observed.

\subsection{Temporal evolution of the particle number operator and 
the expectation value of energy}
\label{sec:bdb}

\begin{figure}
 \centering
   \includegraphics[width=\columnwidth]{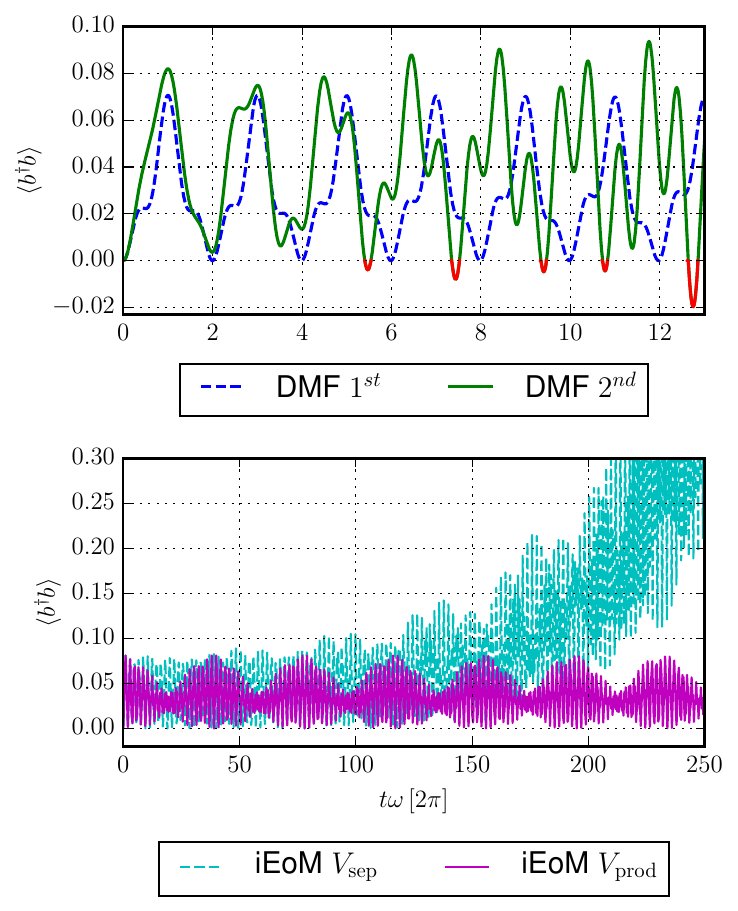}   
   \caption{Results for the temporal evolution of the number operator for
	the parameters $\braket{\sigma_x}(0)=1$, $\omega_0=0.5\omega$, and $g=0.1\omega$. 
	The upper panel displays the DMF results obtained in first and second order. 
	Expectation values with $\braket{b^\dagger b}<0$ occurring in second order are marked in red.
	The lower panel displays the iEoM results based on  $V_{\mathrm{prod}}$ and 
	on $V_{\mathrm{sep}}$ on a large time scale.}
  \label{f:dbd}
  \end{figure} 
	
The main advantage of the iEoM over the DMF is that products of operators with 
their hermitian conjugates will yield non-negative results by construction.
By this argument, the operator basis $V_{\mathrm{sep}}$ comprising $b$ as well as $b^\dagger$
\textit{guarantees} a non-negative temporal evolution of the number operator. 
We observe, however, that the expectation values of the number operator $b^\dagger b$
computed in the operator basis $V_{\mathrm{prod}}$ does not display negative values either.
So these results are physically reasonable. But at present, 
we do not see a compelling mathematical argument why this is so.

In the upper panel of Fig.~\ref{f:dbd} it is evident, that the DMF results in 
second order display unphysical behavior, namely particle densities falling below zero, 
even for short times. These negative expectation values are highlighted in red. 
By contrast, the  expectation value of the number operator remains non-negative for all 
times within the iEoM approach. 

While the expectation value computed using $V_{\mathrm{prod}}$ shows regular oscillations 
and does not exceed a maximum value the expectation value 
computed using $V_{\mathrm{sep}}$ diverges exponentially. 
This must be classified unphysical since the full quantum mechanical dynamics 
is unitary and thus cannot yield diverging results. For instance, the 
expectation value of the total energy has to stay constant.

Computing the expectation value of the energy using the iEoM, it is important to calculate 
the expectation value $\braket{H_\text{I}}=g\braket{\sigma_x\left(b+b^\dagger\right)}$ 
using the evolution of $\sigma_x b$ and $\sigma_x b^\dagger$ which are elements of both
 operator bases.  Using the product of the temporal evolutions of $\sigma_x$ and $b$ or 
$b^\dagger$ computed separately, the argument implicitly assumes that 
$\left[\sigma_i,b^\mu\right]=0$. But this does not hold true for the truncated temporal 
evolutions of the operators except at $t=0$ for trivial reason. Therefore, the  
expectation value $\braket{\sigma_x(t)\left(b(t)+b^\dagger(t)\right)}$ computed from the
product has a non-negligible imaginary part even though the operator is hermitian. 
By contrast, the expectation value computed using $\sigma_x b(t)$ and $\sigma_x b^\dagger(t)$ 
remains real for all times as it has to be.

\begin{figure}
 \centering
   \includegraphics[width=\columnwidth]{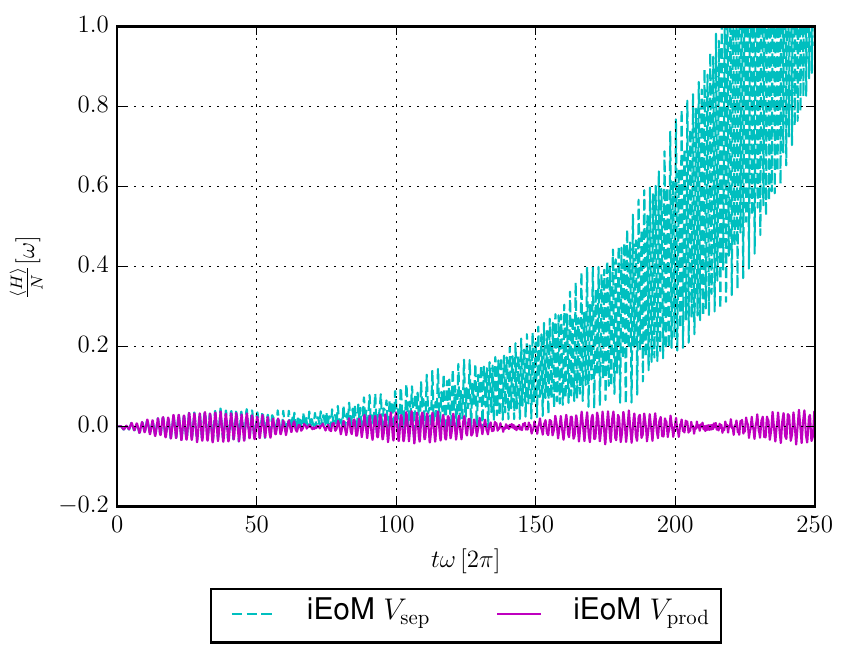} 
   \caption{iEoM results for the temporal evolution of the expectation value of the energy 
	for 	$\braket{\sigma_x}(0)=1$, $\omega_0=0.5\omega$, and $g=0.1\omega$ using the 
	operator bases 	$V_{\mathrm{prod}}$ and $V_{\mathrm{sep}}$.}
  \label{f:Energie_iEoM}
\end{figure}

The temporal evolution of the expectation value of energy is displayed in 
Fig.~\ref{f:Energie_iEoM}. As expected, the evolution computed using 
$V_{\mathrm{sep}}$ diverges. This divergence is solely due to the contribution of the 
number operator to $\braket{H_\text{R}}$, as the temporal evolution of 
$\braket{\sigma_x b}$ is the same for $V_{\mathrm{sep}}$ and for $V_{\mathrm{prod}}$.

Nevertheless, since all calculations are exact in second order in $g$, no matter if 
they are based on $V_{\mathrm{sep}}$ or on $V_{\mathrm{prod}}$, the error of 
$\braket{H_\text{R}}$ is of order $g^3$ or possibly of higher order. 
The plot of the deviation of 
$\braket{H_\text{R}}$ from zero on a double logarithmic scale reveals that for 
small couplings the proportionality $\braket{H_\text{R}}\propto g^4 $ holds true.
This behavior agrees with our expectation because $H_\text{R}$ is 
invariant under the transformation 
\begin{subequations}
\begin{eqnarray}
g&\leftrightarrow& -g\\
b&\leftrightarrow& -b\,,
\end{eqnarray}
\end{subequations}
which implies that the ground state energy and certain other expectation values only 
depend on even powers of $g$. Thus, the accuracy of the energy in order $g^3$ implies the
accuracy in order $g^4$.

Note that $V_{\mathrm{sep}}$ is the operator basis that leads to a significantly better 
agreement with the exact results for the temporal evolution of the spin operators, in spite
of the detrimental long time behavior discernible in Fig.~\ref{f:Energie_iEoM}.
Contrary to the energy computed using $V_{\mathrm{sep}}$, the 
energy using $V_{\mathrm{prod}}$ does not diverge. Here, the expectation value 
oscillates around zero and shows beating behavior with a constant amplitude for all times. 
So again, we observe that the conceptually more suitable operator basis leads to results 
that deviate more from the exact solution for intermediate times.

\section{Conclusions and Outlook}
\label{sec:conclusio}

\subsection{Objective}

We studied the real time behavior in the 
quantum Rabi model (QRM) off equilibrium by two related, but distinct approximate approaches.
The QRM is chosen because it represents a correlated non-trivial model
for which an exact solution for the non-equilibrium dynamics exists 
\cite{braak11,wolf12}.

Both approximate approaches to be compared are based on the Heisenberg equations of motion.
The first approach is the widely used
density-matrix formalism (DMF) which directly focuses on  the time
dependence of expectation values \cite{rossi02,akbar13,krull14,krull16}. The second approach 
is dubbed iterated equations of motion (iEoM) and focuses on the time
evolution of the operators described in a basis of time independent operators
\cite{uhrig09c}.
The motivation is to study how well these approaches work and to
judge their strengths and weaknesses. This will help to employ them 
for other systems more efficiently.

\subsection{Summary}

We found that the DMF applied to the QRM  in second order leads to non-physical results.
The expectation values of the non-negative number operator turn spuriously negative. 
Furthermore, significant deviations from the exact solution occur.

The iEoM approach requires to work with an operator basis chosen beforehand. Different
choices of this basis lead to different dynamic matrices and thus to different results. 
A key advantage of the iEoM approach is that densities stay non-negative by 
construction as it has to be.

Conceptually, the dynamic matrices describing
the operator dynamics in the iEoM should have only real eigenvalues because only
real frequencies guarantee solutions of merely oscillatory character avoiding
exponential divergences. It is such oscillatory behavior which is characteristic
for unitary quantum dynamics. This is ensured formally for hermitian
dynamic matrices. 

For the QRM, an operator basis ($V_\text{sep}$) comprising both $b$ and $b^\dagger$ leads 
to a dynamic matrix with complex eigenvalues. Still, non-negative expectation values for the number operator are guaranteed. The complex eigenvalues, however, imply that 
the expectation values of the number operator show exponential divergence. In addition, unphysical results occur for the expectation values of the spin operators
for longer times.

An alternative operator basis  ($V_\text{prod}$) avoiding the single operators $b$ and $b^\dagger$ but including the product $ b^\dagger b$
leads to a dynamic matrix with only real eigen values. Here, no unphysical results or divergences occur. The number operator remains non-negative, even though this is not 
guaranteed by construction. Therefore, this is the basis of choice on conceptual 
grounds. 

But we observe that the former choice of basis agrees significantly better than
the latter choice in spite of the conceptual assets of the latter. At first sight,
this appears astonishing. At second thought, however, the explanation is that
it is strongly advantageous to include the fundamental operators 
$b$ and $b^\dagger$ in the basis even though their inclusion makes some eigenvalues
of the dynamic matrix complex.

It is a general caveat in dealing with bosonic
systems that the dynamic matrices turn out to be non-hermitian so that
the conceptual issues cannot be avoided. Hence, for short and moderate times
a better agreement can be achieved, but the solutions for long times are spoilt
by the conceptual deficiencies. 

\subsection{Conclusions}

Comparing the DMF and the iEoM we conclude from our results 
that the iEoM is advantageous, at least for the QRM, but also generally.
One general advantage consists in the possibility to improve the approximation
in a systematic way even if no small parameter exists. A larger operator basis
leads to improved results.  Another advantage consists in the preserved
non-negativity of operators.

A caveat of the iEoM applied to the QRM is that we could not find 
large operator bases implying a hermitian dynamic matrix. From our findings,
such a basis should yield good and systematically controlled results.
We tried to identify modified scalar products for the operators to reach
hermitian dynamic matrices, but failed to conceive such scalar products which
are easy to use in practice. 

\subsection{Outlook}

A key problem in the search for better
adapted scalar products is the infinitely large Hilbert space
of bosonic modes. For finite local Hilbert spaces, the situation is 
decisively different and much more promising. Thus, models
consisting only of spins and lattice fermions can be treated by iEoM
starting from operator bases with hermitian dynamic matrices so that 
the approach starts from firm conceptual underpinnings.

The appropriate scalar product between two operators $A$ and $B$ 
reads
\begin{equation}
(A|B) := \frac{1}{d} \text{Tr}(A^\dag B)
\end{equation}
where $d$ is the dimension of the total Hilbert space of the model.
Clearly, this scalar product is well defined on any finite lattice.
The thermodynamic limit of infinite lattices is also defined if the 
operators $A$ and $B$ have a finite spatial support, i.e., they
are defined on a cluster with a finite number of lattice sites.

The key observation ensuring that the dynamic matrix $\uuline{M}$ is
hermitian is the following identity for hermitan Hamiltonians $H$
\begin{subequations}
\begin{eqnarray}
(A|[H,B]) &=& \frac{1}{d} \left[ \text{Tr}(A^\dag HB) - \text{Tr}(A^\dag BH)\right] 
\\
&=&\frac{1}{d} \left[ \text{Tr}((HA)^\dag B) - \text{Tr}((AH)^\dag B)\right] 
\\
&=& ([H,A] |B).
\end{eqnarray}
\end{subequations}
This means that the commutation with $H$, also called
Liouvillean super-operator, is self-adjoint. Thus its
representation as a matrix  with respect to an orthonormal basis 
is hermitian. Therefore, by the above very general argument, we have 
devised a way to avoid exponential divergences and to construct
approximations displaying oscillatory behavior as is generic
in quantum mechanics. This route of iEoM for non-equilibrium
physics shall be explored further in future research.

In parallel, we think that it will be also instructive to
compare the DMF and the iEoM approach to the standard diagrammatic
approach based on Keldysh diagrams \cite{freer06,kamen11}. This is another interesting line of research.

\acknowledgement
We thank Ilya Eremin, Marcus Kollar, Dirk Manske and Andreas Schnyder
 for helpful discussions. We are grateful to Marcus Kollar for the provision of
data. Financial support of the DFG is acknowledged in TRR 160.

\subsection{Author contribution statement}
After G.S.U. had chosen the topic of the study, H.K. performed the first calculations and obtained the first results. This was taken up by M.K., who - with the support of F.K. - did further calculations and obtained the final results,  which were interpreted by G.S.U.. The manuscript has been written by M.K. ($\approx 75$\%) and G.S.U. ($\approx 25$\%) and edited by F.K..

\section{Appendix}
\subsection{Differential Equations of the QRM}
The differential eqations resulting from the Heisenberg equation of motion for the QRM are given by  
\allowdisplaybreaks
\begin{subequations}
\begin{eqnarray}
\frac{\mathrm{d}}{\mathrm{d}t} b &=& -i\omega b -ig\sigma_x\\
\frac{\mathrm{d}}{\mathrm{d}t} b^\dagger &=& i\omega b^\dagger +ig\sigma_x\\
\frac{\mathrm{d}}{\mathrm{d}t} \sigma_x &=&-\omega_0 \sigma_y\\
\frac{\mathrm{d}}{\mathrm{d}t} \sigma_y &=&\omega_0\sigma_x -2g\sigma_z b -2g \sigma_z b^\dagger\\
\frac{\mathrm{d}}{\mathrm{d}t} \sigma_z &=& 2g\sigma_y b +2g\sigma_y b^\dagger\\
\frac{\mathrm{d}}{\mathrm{d}t} \sigma_x b &=& -i\omega \sigma_x b -\omega_0 \sigma_y b -gi\mathbbm{1}  \\
\frac{\mathrm{d}}{\mathrm{d}t} \sigma_x b^\dagger &=& i\omega \sigma_x b^\dagger -\omega_0 \sigma_y b^\dagger +gi\mathbbm{1}  \\\nonumber
\frac{\mathrm{d}}{\mathrm{d}t} \sigma_y b &=& -i\omega\sigma_y b +\omega_0 \sigma_x b -g\sigma_z \\
&&-2g\sigma_z bb -2g\sigma_z b^\dagger b\\\nonumber
\frac{\mathrm{d}}{\mathrm{d}t} \sigma_y b^\dagger &=& i\omega\sigma_y b^\dagger +\omega_0 \sigma_x b^\dagger -g\sigma_z \\
&&-2g\sigma_z b^\dagger b^\dagger -2g\sigma_z b^\dagger b\\
\frac{\mathrm{d}}{\mathrm{d}t} \sigma_z b &=& -i\omega\sigma_z b +g \sigma_y +2g \sigma_y b b + 2g\sigma_y b^\dagger b\\
\frac{\mathrm{d}}{\mathrm{d}t} \sigma_z b^\dagger &=& i\omega\sigma_z b^\dagger +g \sigma_y\nonumber\\
&& +2g \sigma_y b^\dagger b^\dagger + 2g\sigma_y b^\dagger b\\
\frac{\mathrm{d}}{\mathrm{d}t} bb&=&-2 i \omega bb -2gi \sigma_x b\\
\frac{\mathrm{d}}{\mathrm{d}t} b^\dagger b^\dagger &=&-2 i \omega b^\dagger b^\dagger  -2gi \sigma_x b^\dagger\\
\frac{\mathrm{d}}{\mathrm{d}t} b^\dagger b&=& ig \sigma_x b -ig \sigma_x b^\dagger\\
\frac{\mathrm{d}}{\mathrm{d}t} \sigma_x bb &=&-\omega_0 \sigma_y bb -2i\omega\sigma_x bb -2ig b\\
\frac{\mathrm{d}}{\mathrm{d}t} \sigma_x b^\dagger b^\dagger  &=&-\omega_0 \sigma_y b^\dagger b^\dagger  +2i\omega\sigma_x b^\dagger b^\dagger  +2ig b^\dagger \\
\frac{\mathrm{d}}{\mathrm{d}t} \sigma_x b^\dagger b  &=&-\omega_0 \sigma_y b^\dagger b  +gi b -gi b^\dagger \\\nonumber
\frac{\mathrm{d}}{\mathrm{d}t} \sigma_y b b  &=& \omega_0 \sigma_x bb -2i\omega \sigma_y b b - 2g\sigma_zb\\
&& -2g\left(\sigma_z bbb+\sigma_z b^\dagger bb\right)\\\nonumber
\frac{\mathrm{d}}{\mathrm{d}t} \sigma_y b^\dagger b^\dagger  &=& \omega_0 \sigma_x b^\dagger b^\dagger  +2i\omega \sigma_y b^\dagger b^\dagger - 2g\sigma_zb^\dagger\\
&& -2g\left(\sigma_z b^\dagger b^\dagger  b^\dagger +\sigma_z b^\dagger b^\dagger b\right)\\
\frac{\mathrm{d}}{\mathrm{d}t} \sigma_y b^\dagger b  &=& \omega_0 \sigma_x b^\dagger b -g\sigma_z b -g\sigma_z b^\dagger\nonumber\\
&& -2g\left(\sigma_z b^\dagger b^\dagger  b +\sigma_z b^\dagger b b\right)\\
\frac{\mathrm{d}}{\mathrm{d}t} \sigma_z b b  &=& -2i\omega \sigma_z bb +2g \sigma_y b\nonumber\\
&& + 2g\left( \sigma_y b b b +\sigma_y b^\dagger b b \right) \\
\frac{\mathrm{d}}{\mathrm{d}t} \sigma_z b^\dagger b^\dagger  &=& 2i\omega \sigma_z b^\dagger b^\dagger +2g \sigma_y b^\dagger\nonumber\\
&& + 2g\left( \sigma_y b^\dagger b^\dagger b^\dagger +\sigma_y b^\dagger b^\dagger b \right) \\
\frac{\mathrm{d}}{\mathrm{d}t} \sigma_z b^\dagger b  &=& g\sigma_y b + g\sigma_y b^\dagger\nonumber\\
&& + 2g\left( \sigma_y b^\dagger b^\dagger b +\sigma_y b^\dagger b b \right)
\end{eqnarray}
\end{subequations}
\subsection{Temporal evolution of the cumulants}

\begin{subequations}
\begin{eqnarray}
 \frac{\mathrm{d}}{\mathrm{d}t} \braket{b}^c &=& -i\omega\braket{b} -ig\braket{\sigma_x}\\
 \frac{\mathrm{d}}{\mathrm{d}t} \braket{b^\dagger}^c &=& i\omega\braket{b^\dagger} +ig\braket{\sigma_x}\\
 \frac{\mathrm{d}}{\mathrm{d}t} \braket{\sigma_x}^c &=& -\omega_0\braket{\sigma_y}\\
 \frac{\mathrm{d}}{\mathrm{d}t} \braket{\sigma_y}^c &=& \omega_0\braket{\sigma_x}
-2g\left( \braket{\sigma_z b^\dagger }^c +\braket{\sigma_z}\braket{b^\dagger}\right)\nonumber\\
&&-2g\left(\braket{\sigma_z b }^c +\braket{\sigma_z}\braket{b}\right) \\
 \frac{\mathrm{d}}{\mathrm{d}t} \braket{\sigma_z}^c &=& +2g\left( \braket{\sigma_y b^\dagger }^c +\braket{\sigma_y}\braket{b^\dagger} \right)\nonumber\\
 &&+2g\left(\braket{\sigma_y b }^c +\braket{\sigma_y}\braket{b}\right)\\
\frac{\mathrm{d}}{\mathrm{d}t} \braket{b b}^c &=& -2i\omega \braket{b b }^c -2ig \braket{\sigma_x b}^c\\
\frac{\mathrm{d}}{\mathrm{d}t} \braket{b^\dagger b^\dagger}^c &=& 2i\omega \braket{b^\dagger b^\dagger }^c +2ig \braket{\sigma_x b^\dagger}^c\\
\frac{\mathrm{d}}{\mathrm{d}t} \braket{b^\dagger b}^c &=& -ig \braket{\sigma_x b^\dagger}^c+ig \braket{\sigma_x b}^c\\
\frac{\mathrm{d}}{\mathrm{d}t} \braket{\sigma_x b}^c &=& -i\omega\braket{\sigma_x b}^c-\omega_0 \braket{\sigma_y b}^c \nonumber\\
&&+gi \braket{\sigma_x} \braket{\sigma_x} -gi\\
\frac{\mathrm{d}}{\mathrm{d}t} \braket{\sigma_x b^\dagger}^c &=& i\omega\braket{\sigma_x b^\dagger}^c-\omega_0 \braket{\sigma_y b^\dagger}^c\nonumber\\
&& -gi \braket{\sigma_x} \braket{\sigma_x} +gi\\\nonumber
\frac{\mathrm{d}}{\mathrm{d}t} \braket{\sigma_y b}^c &=& ig\braket{\sigma_x}\braket{\sigma_y}-i\omega \braket{\sigma_y b}^c\nonumber\\\nonumber
&& -g\braket{\sigma_z}+\omega_0\braket{\sigma_x b}^c \\\nonumber
&&{+2g\left(\braket{\sigma_z bb}^c +\braket{\sigma_z b^\dagger b}^c\right)} \\\nonumber
&&+2g\left(
\braket{\sigma_z b}^c \braket{b}+\braket{\sigma_z b}^c \braket{b^\dagger}\right)\\
&&+2g\left( \braket{b b}^c \braket{\sigma_z}+\braket{b^\dagger b}^c \braket{\sigma_z} \right)\\\nonumber
\frac{\mathrm{d}}{\mathrm{d}t} \braket{\sigma_y b^\dagger }^c &=& -ig\braket{\sigma_x}\braket{\sigma_y}+i\omega \braket{\sigma_y b^\dagger }^c\\\nonumber
&& -g\braket{\sigma_z}+\omega_0\braket{\sigma_x b^\dagger }^c \\\nonumber
&&{+2g\left(\braket{\sigma_z b^\dagger b^\dagger }^c +\braket{\sigma_z b^\dagger b}^c \right)}\\\nonumber
&&+2g\left(\braket{\sigma_z b^\dagger }^c \braket{b^\dagger }+\braket{\sigma_z b^\dagger }^c \braket{b}\right)\\
&&+2g\left( +\braket{b^\dagger  b^\dagger }^c \braket{\sigma_z}+\braket{b^\dagger b}^c \braket{\sigma_z} \right)\\\nonumber
\frac{\mathrm{d}}{\mathrm{d}t} \braket{\sigma_z b}^c &=& ig\braket{\sigma_x}\braket{\sigma_z}-i\omega \braket{\sigma_z b}^c +g\braket{\sigma_y}\\\nonumber
&&+2g\left(\braket{\sigma_y bb}^c +\braket{\sigma_y b^\dagger b}^c\right)\\\nonumber
&&+2g\left(\braket{\sigma_y b}^c \braket{b}+\braket{\sigma_y b\braket{b^\dagger}}^c\right)\\
&&+2g\left( \braket{b b}^c \braket{\sigma_y}+\braket{b^\dagger b}^c \braket{\sigma_y} \right)\\\nonumber
\frac{\mathrm{d}}{\mathrm{d}t} \braket{\sigma_z b^\dagger }^c &=& -ig\braket{\sigma_x}\braket{\sigma_z}+i\omega \braket{\sigma_z b^\dagger }^c +g\braket{\sigma_y}\\\nonumber
&&{+2g\left(\braket{\sigma_y b^\dagger b^\dagger }^c +\braket{\sigma_y b^\dagger b}^c\right)} \\\nonumber
&&+2g\left(\braket{\sigma_y b^\dagger }^c \braket{b^\dagger }+\braket{\sigma_y b^\dagger }^c \braket{b}\right)\\
&&+2g\left( +\braket{b^\dagger  b^\dagger }^c \braket{\sigma_y}+\braket{b^\dagger b}^c \braket{\sigma_y} \right)
\end{eqnarray}
\end{subequations}
\subsection{Matrices for the IEoM}
To illustrate the vector $\vec{V}$ that contains all operators that are elements of the chosen operator basis and the corresponding matrix $\uuline{M}$ that satisfies Eq.~\eqref{eqn:matrixdgl1},
both the vector and the matrix are divided into smaller submatrices. Eq.~\eqref{eqn:matrixdgl1} then is given by 
\begin{equation}
\frac{\mathrm{d}}{\mathrm{d}t}
\begin{pmatrix}
\vec{V_0} \\
\vec{V_1} \\
\vec{V_1} \\
\vec{V_3} \\
\vec{V_4} 
\end{pmatrix}=i
\left( \begin{array}{c|c c c c }
\uuline{A_{00}} & \uuline{A_{01}} & \uuline{A_{02}} & \uuline{A_{03}} & \uuline{A_{04}} \\
    \hline
\uuline{A_{10}} & \uuline{C_{11}} & \uuline{C_{12}} & \uuline{C_{13}} & \uuline{C_{14}} \\
\uuline{A_{20}} & \uuline{C_{21}} & \uuline{C_{22}} & \uuline{C_{23}} & \uuline{C_{24}} \\
\uuline{A_{30}} & \uuline{C_{31}} & \uuline{C_{32}} & \uuline{C_{33}} & \uuline{C_{34}} \\
\uuline{A_{40}} & \uuline{C_{41}} & \uuline{C_{42}} & \uuline{C_{43}} & \uuline{C_{44}} \\
\end{array}\right)
\begin{pmatrix}
\vec{V_0} \\
\vec{V_1} \\
\vec{V_2} \\
\vec{V_3} \\
\vec{V_4} 
\end{pmatrix}\,,
\end{equation}
where the vectors $\vec{V}_n$ for $n> 0$ and the matrices $\uuline{C_{nm}}$ are the same in both $V_{\mathrm{prod}}$ and $V_{\mathrm{sep}}$. They are given by
\begin{subequations}
\begin{eqnarray}
\vec{V}_1=
\begin{pmatrix}
\sigma_x\\
\sigma_y\\
\sigma_z
\end{pmatrix},&\,&
\vec{V}_2=
\begin{pmatrix}
\sqrt{2}\sigma_x b\\
\sqrt{2}\sigma_x b^\dagger\\
\sqrt{2}\sigma_y b\\
\sqrt{2}\sigma_y b^\dagger\\
\sqrt{2}\sigma_z b\\
\sqrt{2}\sigma_z b^\dagger\\
\end{pmatrix}\\
\vec{V}_4=
\begin{pmatrix}
\sqrt{2}\sigma_x b b\\
\sqrt{2}\sigma_x b^\dagger b^\dagger\\
\sqrt{2}\sigma_y b b\\
\sqrt{2}\sigma_y b^\dagger b^\dagger\\
\sqrt{2}\sigma_z b b\\
\sqrt{2}\sigma_z b^\dagger b^\dagger\\
\end{pmatrix},&\,&
\vec{V}_5=
\begin{pmatrix}
2\sigma_x b^\dagger b\\
2\sigma_y b^\dagger b\\
2\sigma_z b^\dagger b\\
\end{pmatrix}\,.
\end{eqnarray} 
\end{subequations}
The matrices $\uuline{C_{nm}}$ satisfy 
\begin{eqnarray}
\uuline{C_\mathrm{nm}}&=&\left(\uuline{C_\mathrm{mn}}\right)^\dagger
\end{eqnarray}
and are given by
\begin{subequations}
\begin{eqnarray}
\uuline{C_{11}}=\uuline{C_{44}}&=&\begin{pmatrix}
0 & i\omega_0 & 0\\
-i\omega_0 & 0 & 0\\
0 & 0 & 0
\end{pmatrix}
\\
\uuline{C_{22}}&=&\begin{pmatrix}
-\omega & 0 & i\omega_0 & 0 & 0 & 0\\
0 & \omega & 0 & \-i\omega_0 & 0 & 0 \\
-i\omega_0 & 0 & -\omega & 0 & 0 & 0\\
0 & -i\omega_0 & 0 & \omega & 0 & 0\\
0 & 0 & 0 & 0 & -\omega & 0\\
0 & 0 & 0 & 0 & 0 & \omega
\end{pmatrix}\\
\uuline{C_{33}}&=&\begin{pmatrix}
-2\omega & 0 & i\omega_0 & 0 & 0 & 0\\
0 & 2\omega & 0 & \-i\omega_0 & 0 & 0 \\
-i\omega_0 & 0 & -2\omega & 0 & 0 & 0\\
0 & -i\omega_0 & 0 & 2\omega & 0 & 0\\
0 & 0 & 0 & 0 & -2\omega & 0\\
0 & 0 & 0 & 0 & 0 & 2\omega
\end{pmatrix}
\\
\uuline{C_{12}}&=&\begin{pmatrix}
0 & 0 & 0 & 0 & 0 & 0\\
0 & 0 & 0 & 0 & \sqrt{2}ig & \sqrt{2}i g\\
0 & 0 & 0 & -\sqrt{2}ig & -\sqrt{2}i g & 0
\end{pmatrix}
\\
\uuline{C_{23}}&=&\begin{pmatrix}
0 & 0 & 0 & 0 & 0 & 0\\
0 & 0 & 0 & 0 & 0 & 0\\
0 & 0 & 0 & 0 & 2ig & 0\\
0 & 0 & 0 & 0 & 0 & 2ig\\
0 & 0 & -2ig & 0 & 0 & 0\\
0 & 0 & 0 & -2ig & 0 & 0
\end{pmatrix}\\
\uuline{C_{24}}&=&
\begin{pmatrix}
0 & 0 & 0\\
0 & 0 & 0\\
0 & 0 & \sqrt{2}ig\\
0 & 0 & \sqrt{2}ig\\
0 & -\sqrt{2}ig & 0\\
0 & -\sqrt{2}ig & 0
\end{pmatrix}\\
\uuline{C_{13}}&=&
\uuline{C_{34}}=\uuline{0}\,.
\end{eqnarray} 
\end{subequations}
For $V_{\mathrm{prod}}$, with
\begin{eqnarray}
\begin{subequations}
\vec{V}_{0,\mathrm{prod}}=
\begin{pmatrix}
\mathbbm{1}\\
\sqrt{2}b^\dagger b
\end{pmatrix}
\end{subequations}
\end{eqnarray}
the matrices $\uuline{A_{nm,\text{prod}}}$ are given by
\begin{subequations}
\begin{eqnarray}
\uuline{A_{02,\text{prod}}}&=&
\begin{pmatrix}
0&0&0&0&0&0\\
g&-g&0&0&0&0
\end{pmatrix}\\
\uuline{A_{20,\text{prod}}}&=&
\begin{pmatrix}
-\sqrt{2}g&0\\
\sqrt{2}g&0\\
0&0\\
0&0\\
0&0\\
0&0
\end{pmatrix}\\
\uuline{A_{00,\text{prod}}}&=&\uuline{A_{01,\text{prod}}}=\uuline{A_{03,\text{prod}}}= \uuline{A_{04,\text{prod}}}=\uuline{0}\\
\uuline{A_{10,\text{prod}}}&=&\uuline{A_{30,\text{prod}}}=\uuline{A_{40,\text{prod}}}= \uuline{0}\,.
\end{eqnarray}
\end{subequations}
For $V_{\mathrm{sep}}$, with
\begin{eqnarray}
\begin{subequations}
\vec{V}_{0,\mathrm{sep}}=
\begin{pmatrix}
\mathbbm{1}\\
b\\
b^\dagger
\end{pmatrix}
\end{subequations}
\end{eqnarray}
the matrices $\uuline{A_{nm,\text{sep}}}$ are given by
\begin{subequations}
\begin{eqnarray}
\uuline{A_{00,\text{sep}}}&=&
\begin{pmatrix}
0&0&0\\
0&-\omega & 0\\
0&0&\omega
\end{pmatrix}\\
\uuline{A_{01,\text{sep}}}&=&
\begin{pmatrix}
0&0&0\\
-g&0&0\\
g&0&0
\end{pmatrix}\\
\uuline{A_{20,\text{sep}}}&=&
\begin{pmatrix}
-\sqrt{2}g&0&0\\
\sqrt{2}g&0&0\\
0&0&0\\
0&0&0\\
0&0&0\\
0&0&0
\end{pmatrix}\\
\uuline{A_{30,\text{sep}}}&=&
\begin{pmatrix}
0&-2\sqrt{2}g&0\\
0&0&2\sqrt{2}g\\
0&0&0\\
0&0&0\\
0&0&0\\
0&0&0
\end{pmatrix}\\
\uuline{A_{40,\text{sep}}}&=&
\begin{pmatrix}
0&2g&-2g\\
0&0&0\\
0&0&0
\end{pmatrix}
\\
\uuline{A_{02,\text{sep}}}&=&\uuline{A_{03,\text{sep}}}= \uuline{A_{04,\text{sep}}}= \uuline{A_{10,\text{sep}}}=\uuline{0}\,.
\end{eqnarray}
\end{subequations}
Hence the matrix $\uuline{M}_\text{prod}$ corresponding to $V_\text{prod}$ is given by 
\begin{equation}
\uuline{M}_\text{prod}=\left( \begin{array}{c|c c c c }
\uuline{0} & \uuline{0} & \uuline{A_{02,\text{prod}}} & \uuline{0} & \uuline{0} \\
    \hline
\uuline{0} & \uuline{C_{11}} & \uuline{C_{12}} & \uuline{0} & \uuline{0} \\
\uuline{A_{20,\text{prod}}} & \left(\uuline{C_{12}} \right)^\dagger& \uuline{C_{22}} & \uuline{C_{23}} & \uuline{C_{24}} \\
\uuline{0} & \uuline{0} & \left(\uuline{C_{32}}\right)^\dagger & \uuline{C_{33}} & \uuline{0} \\
\uuline{0} & 0 & \left(\uuline{C_{24}}\right)^\dagger & \uuline{0} & \uuline{C_{44}} \\
\end{array}\right)\,,
\end{equation} 
and the matrix $\uuline{M}_\text{sep}$ corresponding to $V_\text{sep}$ is given by 
\begin{equation}
\uuline{M}_\text{sep}=\left( \begin{array}{c|c c c c }
 \uuline{A_{00,\text{sep}}} &  \uuline{A_{01,\text{sep}}} &  \uuline{0} & \uuline{0} & \uuline{0} \\
    \hline
\uuline{0} & \uuline{C_{11}} & \uuline{C_{12}} & \uuline{0} & \uuline{0} \\
\uuline{A_{20,\text{prod}}} & \left(\uuline{C_{12}} \right)^\dagger& \uuline{C_{22}} & \uuline{C_{23}} & \uuline{C_{24}} \\
 \uuline{A_{30,\text{sep}}} & \uuline{0} & \left(\uuline{C_{32}}\right)^\dagger & \uuline{C_{33}} & \uuline{0} \\
 \uuline{A_{40,\text{sep}}} & 0 & \left(\uuline{C_{24}}\right)^\dagger & \uuline{0} & \uuline{C_{44}} \\
\end{array}\right)\,.
\end{equation} 
The time evolution of the respective prefactors $\uuline{\gamma}_\text{prod}$ and $\uuline{\gamma}_\text{sep}$ is calculated using 
\begin{subequations}
\begin{eqnarray}
\frac{\mathrm{d}}{\mathrm{d}t}\uuline{\gamma}_\text{prod}&=&i\uuline{\gamma}_\text{prod}\uuline{M}_\text{prod}
\\
\frac{\mathrm{d}}{\mathrm{d}t}\uuline{\gamma}_\text{sep}&=&i\uuline{\gamma}_\text{sep}\uuline{M}_\text{sep}\,.
\end{eqnarray}
\end{subequations}

 \bibliographystyle{epj}
 \bibliography{dmf-ieom-v2}

\begin{thebibliography}{89}

\bibitem{grein02b}
M.~Greiner, O.~Mandel, T.~Esslinger, T.W. H\"ansch, I.~Bloch, Nature
  \textbf{419}, 51 (2002)

\bibitem{kinos06}
T.~Kinoshita, T.~Wenger, D.S. Weiss, Nature \textbf{440}, 900 (2006)

\bibitem{lewen07}
M.~Lewenstein, A.~Sanpera, V.~Ahufinger, B.~Damski, A.~Sen(De), U.~Sen, Adv.
  Phys. \textbf{56}, 243 (2007)

\bibitem{bloch08}
I.~Bloch, J.~Dalibard, W.~Zwerger, Rev. Mod. Phys. \textbf{80}, 885 (2008)

\bibitem{essli10}
T.~Esslinger, Ann. Rev. Condens. Matter Phys. \textbf{1}, 129 (2010)

\bibitem{trotz12}
S.~Trotzky, Y.A. Chen, A.~Flesch, I.P. McCulloch, U.~Schollw\"ock, J.~Eisert,
  I.~Bloch, Nature Phys. \textbf{8}, 325 (2012)

\bibitem{ronzh13}
J.P. Ronzheimer, M.~Schreiber, S.~Braun, S.S. Hodgman, S.~Langer, I.P.
  McCulloch, F.~Heidrich-Meisner, I.~Bloch, U.~Schneider, Phys. Rev. Lett.
  \textbf{110}, 205301 (2013)

\bibitem{stroh07}
N.~Strohmaier, Y.~Takasu, K.~G\"unter, R.~J\"ordens, M.~K\"ohl, H.~Moritz,
  T.~Esslinger, Phys. Rev. Lett. \textbf{99}, 220601 (2007)

\bibitem{schne08}
U.~Schneider, L.~Hackerm\"uller, S.~Will, T.~Best, I.~Bloch, T.A. Costi, R.W.
  Helmes, D.~Rasch, A.~Rosch, Science \textbf{322}, 1520 (2008)

\bibitem{stroh10}
N.~Strohmaier, R.J. D.~Greif, L.~Tarruell, H.~Moritz, T.~Esslinger,
  R.~Sensarma, D.~Pekker, E.~Altman, E.~Demler, Phys. Rev. Lett. \textbf{104},
  080401 (2010)

\bibitem{schne12}
U.~Schneider, L.~Hackerm\"uller, J.P. Ronzheimer, S.~Will, S.~Braun, T.~Best,
  I.~Bloch, E.~Demler, S.~Mandt, D.~Rasch et~al., Nature Phys. \textbf{8}, 213
  (2012)

\bibitem{perto14}
D.~Pertot, A.~Sheikhan, E.~Cocchi, L.A. Miller, J.E. Bohn, M.~Koschorreck,
  M.~K\"ohl, C.~Kollath, Phys. Rev. Lett. \textbf{113}, 170403 (2014)

\bibitem{cocch16}
E.~Cocchi, L.A. Miller, J.H. Drewes, M.~Koschorreck, D.~Pertot, F.~Brennecke,
  M.~K\"ohl, Phys. Rev. Lett. \textbf{116}, 175301 (2016)

\bibitem{will15}
S.~Will, D.~Iyer, M.~Rigol, Nature Comm. \textbf{6}, 6009 (2015)

\bibitem{lubas11}
M.~Lubasch, V.~Murg, U.~Schneider, J.I. Cirac, M.C.B. {n}uls, Phys. Rev. Lett.
  \textbf{107}, 165301 (2011)

\bibitem{kosch13}
M.~Koschorreck, D.~Pertot, E.~Vogt, M.~K\"ohl, Nature Phys. \textbf{9}, 405
  (2013)

\bibitem{kraus14}
J.S. Krauser, U.~Ebling, N.~Fl\"aschner, J.~Heinze, K.~Sengstock,
  M.~Lewenstein, A.~Eckardt, C.~Becker, Science \textbf{343}, 157 (2014)

\bibitem{eblin14}
U.~Ebling, J.S. Krauser, N.~Fl\"aschner, K.~Sengstock, C.~Becker,
  M.~Lewenstein, A.~Eckard, Phys. Rev. X \textbf{4}, 021011 (2014)

\bibitem{brown15}
R.C. Brown, R.~Wyllie, S.B. Koller, E.A. Goldschmidt, M.~Foss-Feig, J.V. Porto,
  Science \textbf{348}, 540 (2015)

\bibitem{bohne16}
J.G. Bohnet, B.C. Sawyer, J.W. Britton, M.L. Wall, A.M. Rey, M.~Foss-Feig, J.J.
  Bollinger, Science \textbf{352}, 1297 (2016)

\bibitem{eiser15}
J.~Eisert, M.~Friesdorf, C.~Gogolin, Nature Phys. \textbf{11}, 124 (2015)

\bibitem{cazal06}
M.A. Cazalilla, Phys. Rev. Lett. \textbf{97}, 156403 (2006)

\bibitem{uhrig09c}
G.S. Uhrig, Phys. Rev. A \textbf{80}, 061602(R) (2009)

\bibitem{fiore10}
D.~Fioretto, G.~Mussardo, New J. Phys. \textbf{12}, 055015 (2010)

\bibitem{sabio10}
J.~Sabio, S.~Kehrein, New J. Phys. \textbf{12}, 055008 (2010)

\bibitem{cazal11}
M.A. Cazalilla, R.~Citro, T.~Giamarchi, E.~Orignac, M.~Rigol, Rev. Mod. Phys.
  \textbf{83}, 1405 (2011)

\bibitem{schur12}
D.~Schuricht, F.H.L. Essler, J. Stat. Mech.: Theor. Exp. p. P040717 (2012)

\bibitem{rentr12}
J.~Rentrop, D.~Schuricht, V.~Meden, New J. Phys. \textbf{14}, 075001 (2012)

\bibitem{barth08}
T.~Barthel, U.~Schollw\"ock, Phys. Rev. Lett. \textbf{100}, 100601 (2008)

\bibitem{iucci10}
A.~Iucci, M.A. Cazalilla, New J. Phys. \textbf{12}, 055019 (2010)

\bibitem{calab12a}
P.~Calabrese, F.H.L. Essler, M.~Fagotti, J. Stat. Mech.: Theor. Exp. p. P07016
  (2012)

\bibitem{calab12b}
P.~Calabrese, F.H.L. Essler, M.~Fagotti, J. Stat. Mech.: Theor. Exp. p. P07022
  (2012)

\bibitem{kenne13}
D.M. Kennes, O.~Kashuba, M.~Pletyukhov, H.~Schoeller, V.~Meden, Phys. Rev.
  Lett. \textbf{110}, 100405 (2013)

\bibitem{daley04}
A.J. Daley, C.~Kollath, U.~Schollw\"ock, G.~Vidal, J. Stat. Mech.: Theor. Exp.
  p. P04005 (2004)

\bibitem{white04a}
S.R. White, A.E. Feiguin, Phys. Rev. Lett. \textbf{93}, 076401 (2004)

\bibitem{schol05}
U.~Schollw\"ock, Rev. Mod. Phys. \textbf{77}, 259 (2005)

\bibitem{schol11}
U.~Schollw\"ock, Ann. of Phys. \textbf{326}, 96 (2011)

\bibitem{manma07}
S.R. Manmana, S.~Wessel, R.M. Noack, A.~Muramatsu, Phys. Rev. Lett.
  \textbf{98}, 210405 (2007)

\bibitem{karra12b}
C.~Karrasch, J.~Rentrop, D.~Schuricht, V.~Meden, Phys. Rev. Lett. \textbf{109},
  126406 (2012)

\bibitem{vidma13}
L.~Vidmar, S.~Langer, I.P. McCulloch, U.~Schneider, U.~Schollw\"ock,
  F.~Heidrich-Meisner, Phys. Rev. B \textbf{88}, 235117 (2013)

\bibitem{zalet15}
M.P. Zaletel, R.S.K. Mong, C.~Karrasch, J.E. Moore, F.~Pollmann, Phys. Rev. B
  \textbf{91}, 165112 (2015)

\bibitem{freer06}
J.K. Freericks, V.M. Turkowski, V.~Zlati\'c, Phys. Rev. Lett. \textbf{97},
  266408 (2006)

\bibitem{eckst09}
M.~Eckstein, M.~Kollar, P.~Werner, Phys. Rev. Lett. \textbf{103}, 056403 (2009)

\bibitem{schmi13}
B.~Schmidt, M.R. Bakhtiari, I.~Titvinidze, U.~Schneider, M.~Snoek,
  W.~Hofstetter, Phys. Rev. Lett. \textbf{110}, 075302 (2013)

\bibitem{aoki14}
H.~Aoki, N.~Tsuji, M.~Eckstein, M.~Kollar, T.~Oka, P.~Werner, Rev. Mod. Phys.
  \textbf{86}, 779 (2014)

\bibitem{schir10}
M.~Schir\'o, M.~Fabrizio, Phys. Rev. Lett. \textbf{105}, 076401 (2010)

\bibitem{schir11}
M.~Schir\'o, M.~Fabrizio, Phys. Rev. B \textbf{83}, 165105 (2011)

\bibitem{rigol07}
M.~Rigol, V.~Dunjko, V.~Yurovsky, M.~Olshanii, Phys. Rev. Lett. \textbf{98},
  050405 (2007)

\bibitem{rigol08}
M.~Rigol, V.~Dunjko, M.~Olshanii, Nature \textbf{452}, 854 (2008)

\bibitem{torre13}
E.J. Torres-Herrera, L.F. Santos, Phys. Rev. E \textbf{88}, 042121 (2013)

\bibitem{bonca99}
J.~Bon\v{c}a, S.A. Trugman, I.~Batisti\'c, Phys. Rev. B \textbf{60}, 1633
  (1999)

\bibitem{bonca07}
J.~Bon\v{c}a, S.~Maekawa, T.~Tohyama, Phys. Rev. B \textbf{76}, 035121 (2007)

\bibitem{mierz11a}
M.~Mierzejewski, L.~Vidmar, J.~Bon\v{c}a, P.~Prelov\v{s}ek, Phys. Rev. Lett.
  \textbf{106}, 196401 (2011)

\bibitem{mierz11b}
M.~Mierzejewski, J.~Bon\v{c}a, P.~Prelov\v{s}ek, Phys. Rev. Lett. \textbf{107},
  126601 (2011)

\bibitem{bonca12}
J.~Bon\v{c}a, M.~Mierzejewski, L.~Vidmar, Phys. Rev. Lett. \textbf{109}, 156404
  (2012)

\bibitem{rigol14}
M.~Rigol, Phys. Rev. Lett. \textbf{112}, 170601 (2014)

\bibitem{rapp10}
A.~Rapp, S.~Mandt, A.~Rosch, Phys. Rev. Lett. \textbf{105}, 220405 (2010)

\bibitem{kolla11}
M.~Kollar, F.A. Wolf, M.~Eckstein, Phys. Rev. B \textbf{84}, 054304 (2011)

\bibitem{polko11}
A.~Polkovnikov, K.~Sengupta, A.~Silva, M.~Vengalattore, Rev. Mod. Phys.
  \textbf{83}, 863 (2011)

\bibitem{farib13a}
A.~Faribault, D.~Schuricht, Phys. Rev. Lett. \textbf{110}, 040405 (2013)

\bibitem{farib13b}
A.~Faribault, D.~Schuricht, Phys. Rev. B \textbf{88}, 085323 (2013)

\bibitem{uhrig14a}
G.S. Uhrig, J.~Hackmann, D.~Stanek, J.~Stolze, F.B. Anders, Phys. Rev. B
  \textbf{90}, 060301(R) (2014)

\bibitem{goth12}
F.~Goth, F.F. Assaad, Phys. Rev. B \textbf{85}, 085129 (2012)

\bibitem{lenar13}
Z.~Lenar\v{c}i\v{c}, P.~Prelov\v{s}ek, Phys. Rev. Lett. \textbf{111}, 016401
  (2013)

\bibitem{orus14}
R.~Or\'us, Ann. of Phys. \textbf{349}, 117 (2014)

\bibitem{gober05}
D.~Gobert, C.~Kollath, U.~Schollw\"ock, G.~Sch\"utz, Phys. Rev. E \textbf{71},
  036102 (2005)

\bibitem{rossi02}
F.~Rossi, T.~Kuhn, Rev. Mod. Phys. \textbf{74}, 895 (2002)

\bibitem{matsu12}
R.~Matsunaga, R.~Shimano, Phys. Rev. Lett. \textbf{109}, 187002 (2012)

\bibitem{matsu13b}
R.~Matsunaga, Y.I. Hamada, K.~Makise, Y.~Uzawa, H.~Terai, Z.~Wang, R.~Shimano,
  Phys. Rev. Lett. \textbf{111}, 057002 (2013)

\bibitem{matsu16}
R.~Matsunaga, N.~Tsuji, H.~Fujita, A.~Sugioka, K.~Makise, Y.~Uzawa, H.~Terai,
  Z.~Wang, H.~Aoki, R.~Shimano, Science \textbf{345}, 1145 (2016)

\bibitem{ramea16}
J.D. Rameau, S.~Freutel, A.F. Kemper, M.A. Sentef, J.K. Freericks, I.~Avigo,
  M.~Ligges, L.~Rettig, Y.~Yoshida, H.~Eisaki et~al., Nature Comm. \textbf{7},
  13761 (2016)

\bibitem{perfe06}
L.~Perfetti, P.A. Loukakos, M.~Lisowski, U.~Bovensiepen, H.~Berger,
  S.~Biermann, P.S. Cornaglia, A.~Georges, M.~Wolf, Phys. Rev. Lett.
  \textbf{97}, 067402 (2006)

\bibitem{schmi08e}
F.~Schmitt, P.S. Kirchmann, U.~Bovensiepen, R.G. Moore, L.~Rettig, M.~Krenz,
  J.H. Chu, N.~Ru, L.~Perfetti, D.H. Lu et~al., Science \textbf{321}, 1649
  (2008)

\bibitem{akbar13}
A.~Akbari, A.P. Schnyder, D.~Manske, I.~Eremin, Europhys. Lett. \textbf{101},
  17002 (2013)

\bibitem{krull14}
H.~Krull, D.~Manske, G.S. Uhrig, A.P. Schnyder, Phys. Rev. B \textbf{90},
  014515 (2014)

\bibitem{kempe15}
A.F. Kemper, M.A. Sentef, B.~Moritz, J.K. Freericks, T.P. Devereaux, Phys. Rev.
  B \textbf{92}, 224517 (2015)

\bibitem{krull16}
H.~Krull, N.~Bittner, G.S. Uhrig, D.~Manske, A.P. Schnyder, Nature Comm.
  \textbf{7}, 11921 (2016)

\bibitem{rabi36}
I.I. Rabi, Phys. Rev. Lett. \textbf{49}, 324 (1936)

\bibitem{braak11}
D.~Braak, Phys. Rev. Lett. \textbf{107}, 100401 (2011)

\bibitem{wolf12}
F.A. Wolf, M.~Kollar, D.~Braak, Phys. Rev. A \textbf{85}, 053817 (2012)

\bibitem{kubo62}
R.~Kubo, J. Phys. Soc. Jpn. \textbf{17}, 1100 (1962)

\bibitem{fulde88}
P.~Fulde, J.~Keller, G.~Zwicknagl, Solid State Physics \textbf{41} (1988)

\bibitem{hamer13a}
S.A. Hamerla, G.S. Uhrig, Phys. Rev. B \textbf{87}, 064304 (2013)

\bibitem{hamer13b}
S.A. Hamerla, G.S. Uhrig, New J. Phys. \textbf{15}, 073012 (2013)

\bibitem{hamer14a}
S.A. Hamerla, G.S. Uhrig, Phys. Rev. B \textbf{89}, 104301 (2014)

\bibitem{krull15a}
H.~Krull, \emph{Conductivity of strongly pumped superconductors} (PhD Thesis,
  available at {\verb|http://t1.physik.uni-dortmund.de/uhrig/phd.html|}, TU
  Dortmund, 2015)

\bibitem{radha93}
K.~Radhakrishnan, A.C. Hindmarsh, \emph{Description and use of lsode, the
  livermore solver for ordinary differential equations}, NASA-RP-1327,
  UCRL-ID-113855, Lawrence Livermore National Laboratory (1993)

\bibitem{dorma86}
J.R. Dormand, P.J. Prince, Comp. \mbox{\&} Maths. with Appls. \textbf{12A},
  1007 (1986)

\bibitem{kamen11}
A.~Kamenev, \emph{Field Theory of Non-Equilibrium Systems} (Cambridge
  University Press, Cambridge, UK, 2011)

\end{thebibliography}
%

\end{document}